\documentclass[prb,nobalancelastpage,twocolumn,superscriptaddress,floatfix,longbibliography]{revtex4-1}

\usepackage{hyperref}
\usepackage{braket}
\usepackage{graphicx}
\usepackage{amssymb}
\usepackage{amsmath}
\usepackage{eucal}
\usepackage{color}
\usepackage{bm}
\newcommand{\e}{{\rm e}}
\newcommand{\rmd}{{\rm d}}
\newcommand{\rmi}{{\rm i}}

\newcommand{\half}{{\textstyle{\frac{1}{2}}}}

\newcommand{\quarter}{{\textstyle{\frac{1}{4}}}}

\newcommand{\eps}{\epsilon}

\newcommand{\Gcoupling}{\Gamma}

\newcommand{\ignore}[1]{\relax}

\definecolor{DarkGreen}{rgb}{0,0.7,0}

\definecolor{RED}{rgb}{1,0,0}

\begin{document}

\title{Adiabatic almost-topological pumping of fractional charges in non-interacting quantum dots}

\author{Masahiro Hasegawa}
\affiliation{Institute for Solid State Physics, The University of Tokyo, Kashiwa, Chiba 277-8581, Japan}

\author{\'Etienne Jussiau}
\author{Robert S.~Whitney}
\affiliation{
Laboratoire de Physique et Mod\'elisation des Milieux Condens\'es, 
Universit\'e Grenoble Alpes and CNRS, BP 166, 38042 Grenoble, France.}

\date{May 29, 2019}
\begin{abstract}
We use exact techniques to demonstrate theoretically the pumping of fractional charges in a single-level non-interacting quantum dot, when the dot-reservoir coupling is adiabatically driven from weak to strong coupling. 
The pumped charge averaged over many cycles is quantized at a fraction of an electron per cycle, determined by the ratio of Lamb shift to level-broadening; this ratio is imposed by the reservoir band-structure.  For uniform density of states,  half an electron is pumped per cycle.
We call this  {\it adiabatic almost-topological pumping}, because
the pumping's Berry curvature is sharply peaked in the parameter space. Hence, so long as the pumping contour does not touch the peak, the pumped charge depends only on how many times the contour winds around the peak (up to exponentially small corrections).  However, the topology does not protect against
non-adiabatic corrections, which grow linearly with pump speed.
In one limit the peak becomes a delta-function, so the adiabatic pumping of fractional charges becomes entirely topological. 
Our results show that quantization of the adiabatic pumped charge at a fraction of an electron 
does not require  fractional particles or other exotic physics. 
\end{abstract}
\maketitle
%=====================================================

%%%%%%%%%%%%%%%%%%%%%%%%%%%%
\begin{figure}[b]
\includegraphics[width=0.9\columnwidth]{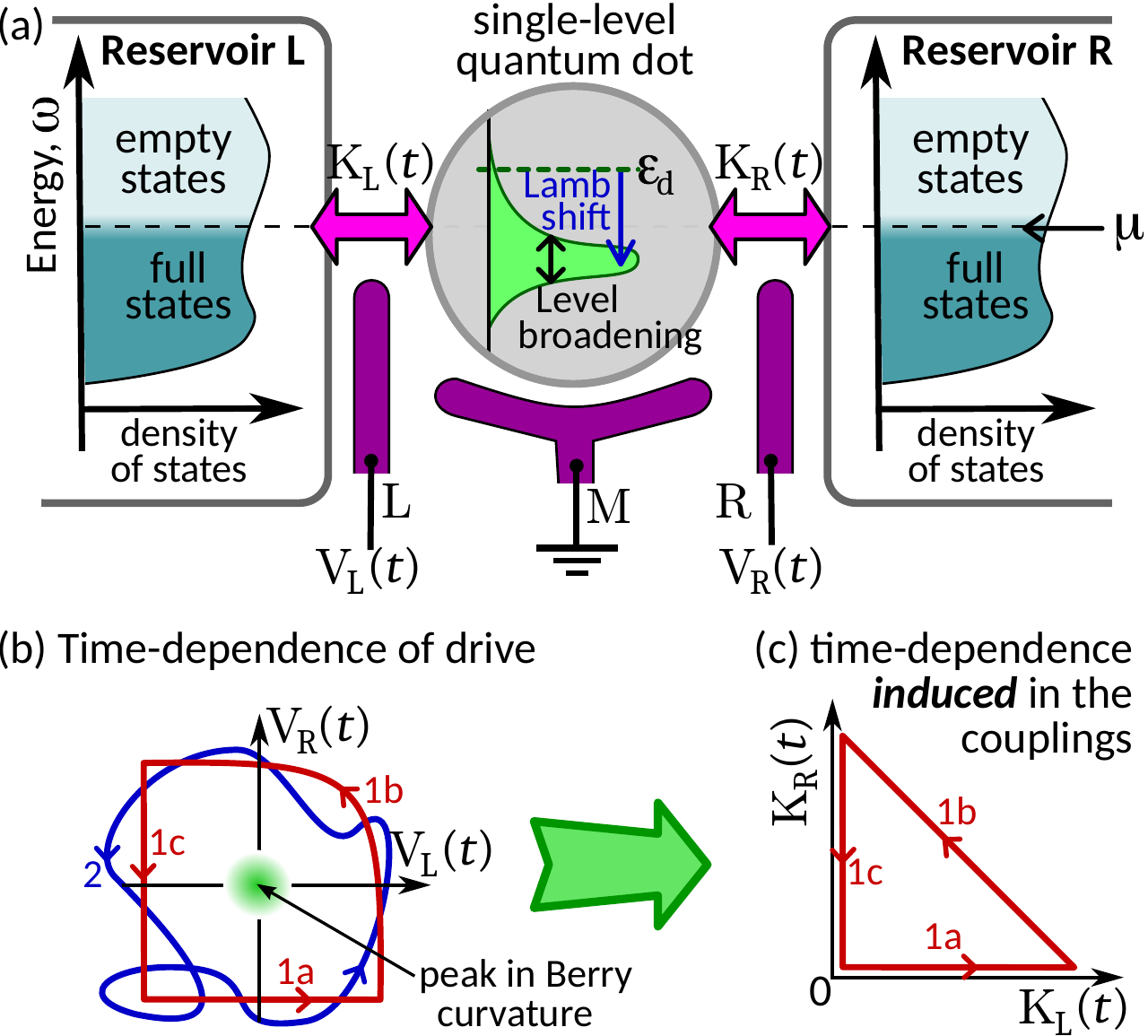}
\caption{\label{Fig:system}
(a) A quantum dot with tunnel-couplings  $K_{\rm L}(t)$ and $K_{\rm R}(t)$  to the reservoirs, 
controlled by gates-voltages, $V_{\rm L}(t)$ and $V_{\rm R}(t)$.
These are slowly varied around the cycle in (b), with gate M  ensuring the dot-level is fixed at energy $\epsilon_{\rm d}$.
Any contour enclosing the Berry curvature peak in (b) without touching it (e.g.\ contours 1 and 2) pumps 
the same fraction of an electron per cycle, up to exponentially small corrections.
The couplings induce level-broadening and a Lamb shift on the dot. 
Since $K_{\rm L}(t)$ and $K_{\rm R}(t)$ depend exponentially on $V_{\rm L,R}(t)$, 
contour 1 in (b) maps to contour 1 in (c).
}
\end{figure}
%%%%%%%%%%%%%%%%%%%%%%%%%

\section{Introduction}
Since the seminal work of Thouless on quantum pumping \cite{Thouless1983May},
there have been many pumping and turnstile protocols discussed in nanoscale systems\cite{Kouwenhoven1991Sep,Pothier1992Dec,Büttiker1994Mar,Aleiner1998Aug,Brouwer1998Oct,Kohler2005Feb,Andergassen2010Jun,Xiao2010Jul,Haupt2013Aug,Pekola2013Oct,Kaestner2015Sep}
and cold atom experiments.\cite{Lohse2015Dec,Nakajima2016Jan}
In recent years there has been great interest in exotic systems which exhibit 
topological pumping of {\it fractional charges}, meaning that any two driving contours with the same topology will drive the same fractional charge.
Such fractional charge pumping has been found in models of 
Coulomb-blockaded quantum dots\cite{Calvo2012dec,Placke2018Aug},
topological insulators\cite{Grusdt2014Nov,Marra2015Mar,Santos2018Jun,Bardyn2019Jan}, 
systems with fractional quantum Hall physics,\cite{Zeng2015Aug,Santos2018Jun}
fermionic gases with short range interactions,\cite{Taddia2017Jun} 
fractional levitons,\cite{Moskalets2016Jul}
and the Bose-Hubbard model.\cite{Gonzalez-Cuadra2019Mar}  
These models have either strong interaction effects or non-trivial topological properties (non-zero Chern numbers, or similar).
This makes us ask if either are necessary; can a non-interacting topologically-trivial system also exhibit 
fractional pumping of a topological nature?

We consider a non-interacting single-level quantum dot at low temperatures. Using the fact that it is an exactly soluble model, we consider adiabatic pumping in this model without approximation (particularly without assuming weak dot-reservoir coupling). 
Our results show the adiabatic pumping of a fraction of an electron in an {\it almost topological} manner. 
That is to say, any pumping contour with the same topology (see Fig.~\ref{Fig:system}b) 
will pump the same fractional charge (up to exponentially small corrections), if the pumping is slow enough to be adiabatic.  However, the topology does not protect against non-adiabatic corrections which go like one over the pumping period. 
This is much less robust than many of the topological pumps mentioned above, in which 
the topology also means that the non-adiabatic corrections decay exponentially with increasing pumping period.

The fractional pumping that we present here occurs when the 
dot-reservoir couplings, $K_{\rm L}$ and $K_{\rm R}$, are adiabatically driven from weak to strong-coupling and back around the pumping cycle, with the dot-level fixed at energy $\eps_{\rm d}$. We take  $\eps_{\rm d}$ to be above the reservoirs' electro-chemical potential, $\mu$.
Here ``strong'' coupling  means that it induces a level-broadening larger than $(\epsilon_{\rm d}-\mu$), so the dot level becomes a resonance that spreads across the electrochemical potential.
The pumped charge is given by the integral over the Berry curvature inside the contour, 
which is sharply peaked and decays exponentially away from the peak.  Formally, the adiabatic pumping would  be topological if this peak was a Dirac $\delta$-function. 
Here the peak has a finite extent, so we refer to the pumping as {\it almost topological}, because it depends only on how many times the contour winds around the peak  --- up to exponentially small corrections --- for any pumping contour that does not impinge on the peak.  
Half an electron is pumped per cycle, if the reservoirs have a uniform density of states (and so impose no Lamb shift of the quantum dot).  However, in general the fraction of an electron pumped per cycle  (between zero and one) is given by the ratio of the Lamb shift imposed by the reservoirs to the level-broadening. This ratio is entirely determined by the reservoirs' density of states, which is imposed by their band-structure.

Earlier works on pumping of dot-reservoir coupling  --- 
with direct driving of the dot-level,\cite{Wohlman02}
a Lamb shift induced by the reservoir band-structure,\cite{Wohlman02,Kashcheyevs2004May} 
Coulomb blockade effects, \cite{Splettstoesser2005Dec,Splettstoesser2006Aug}
or non-adiabatic driving\cite{Battista2011Mar} 
--- did not investigate large level-broadening, and so did not find the quantized pumping of fractional charges. 

Note that we consider the {\it average} charge per cycle. 
There are no fractionally charged quasi-particles in our non-interacting system,
so we expect that there is a certain probability that $n$ electrons are pumped (for integer $n=0,\pm1,\pm2, \cdots$) in any given cycle.  Yet these probabilities are such that the average over many cycles will reveal itself as a fraction per cycle.
Hence the observation of a topological fractional average charge per cycle
in adiabatic pumping does not require the existence of fractionally charged quasi-particles, or other exotic physics. 

It is not yet clear to us if there is a connection to the fractional charges recently discussed in Ref.~[\onlinecite{Riwar2018Nov}].

\subsection{Organisation  of this work}
Sec.~\ref{sec:model} introduces our model Hamiltonian, and 
Sec.~\ref{sec:wideband-explanation} outlines our main result about adiabatic almost-topological pumping of fractional charges.
Sec.~\ref{sec:scattering} shows this is half an electron per cycle for readers familiar with scattering theory (others can skip this section).
Sec.~\ref{sec:dot-occupation} explains that the pumping 
is not simply related to as changes in the dot occupation.
Sec.~\ref{sec:Keldysh} and \ref{sec:general-results} use the Keldysh formalism to get our main result,
Eq.~(\ref{Eq:Pi-versus-Vs}).
Sec.~\ref{sec:ad_condition_be} discusses the non-adiabatic corrections.
Sec.~\ref{sec:conclusion} gives our conclusions.

\section{Model Hamiltonian}
\label{sec:model}
We consider a non-interacting single-level quantum dot connected to two electron reservoirs with time-dependent couplings,
described by the Hamiltonian,
\begin{align}
	H &= \epsilon_{\rm d} d^{\dagger} d + \sum_{i,k} \left[ \epsilon_{k} c^{\dagger}_{ik} c_{ik}  
	+ \gamma_i(t) \big(d^{\dagger} c_{ik} +  c^\dagger_{ik}d \big) \right]. 
	\label{Eq:H}
\end{align}
often called the Fano-Anderson model.\cite{Fano1961,Anderson1961}
Here, $d^{\dagger}$ and $d$  are creation and annihilation operator of the dot state, which has energy $\epsilon_{\rm d}$,
while $c^{\dagger}_{ik}$ and  $c_{ik}$ are those for the state with wavenumber $k$ and energy $\epsilon_k$  in the reservoir $i=L,R$. The tunnel-coupling between the system and the mode $k$ in reservoir $i$ is $\gamma_i(t)$, which is taken to vary slowly with time.
This model neglects electron-electron interactions on the dot; 
the simplest experimental implementation is discussed in Sec.~\ref{sec:requirements}.
The fact this model is quadratic in the creation and annihilation operators means that it is exactly soluble. 
As a result, we will get its adiabatic pumping properties without making any approximations 
(in particular, we will not need to assume weak dot-reservoir coupling).

We take the reservoirs to have a continuum of states, and assume they both have the same density of states $\rho(\omega)$. In general, this density of states may have energy ($\omega$) dependence, band-gaps, etc.  The system's coupling to each reservoir is described in terms of the time-dependent function 
\begin{eqnarray}
\Gcoupling_i(\omega,t) &=& K_i(t) \,\rho(\omega) 
\label{Eq:Gcoupling_i}
\end{eqnarray}
where the coupling parameter $K_i(t)=  |\gamma_i(t)|^2$.
A second crucial quantity for the physics of this model
 is 
\begin{eqnarray}
\Lambda_i (\omega,t) &=& K_i(t) \ \,P\hskip -3.5mm \int {\rm d} \varepsilon \,{ \rho (\varepsilon) \over \omega - \varepsilon} ,
\label{Eq:Lambda_i}
\end{eqnarray}
where the integral is the principal value. 
For compactness of what follows, we also define 
\begin{eqnarray}
\Gcoupling(\omega,t) &=& \Gcoupling_{\rm L}(\omega,t) +  \Gcoupling_{\rm R}(\omega,t)
\label{Eq:Gcoupling}
\\
\Lambda(\omega,t) &=& \Lambda_{\rm L}(\omega,t) +  \Lambda_{\rm R}(\omega,t)
\label{Eq:Lambda}
\end{eqnarray}
We refer to $\Gcoupling_i(\omega,t)$ as {\it level-broadening},
and to $\Lambda_i(\omega,t)$ as a {\it Lamb shift}.
This is a slight abuse of terminology, but it is justified by the dot's local density of states\cite{Fano1961,Anderson1961} being
$\Gcoupling(\omega) \big/\big[\big(\omega -\epsilon_{\rm d} - \Lambda(\omega)\big)^2 + \Gcoupling^2(\omega)\big]$.
So if $\Gcoupling$ and $\Lambda$ are $\omega$-independent, then they are the level-broadening and Lamb shift, respectively.  We simply keep this terminology for cases where 
$\Gcoupling$ and $\Lambda$ have an $\omega$-dependence.

In what follows, our results will be simplest if $K_i$ is written in terms of 
the dimensionless coupling $X_i$,
which measures the level-broadening in units of the distance of the dot level from the electrochemical potential;
\begin{eqnarray} 
X_i={\rho(\mu) \,K_i \over 2(\epsilon_{\rm d}-\mu)} \ \ \hbox{ for } i=L,R,
\label{Eq:rescale}
\end{eqnarray}
where $\rho(\mu)$ is the density of states at the electrochemical potential, and the factor of two makes formulas compact.

We drive the gate-voltages $V_i$, not  the couplings $K_i$, so we need a relation between them.
Typically, the dot is coupled to reservoir $i$ through tunnel-barriers of height $E_i$ and width $L_i$, so  $K_i \sim\exp [-\kappa_i]$ with $\kappa_i = \sqrt {2mE_i(V_i)}\,L_i(V_i)/\hbar$.  
For large $L_i$ and $E_i$,  even small changes in $V_i$ will make large percentage changes in $X_i$,
so we can expand up to linear order in $V_i$ about $X_i=1$.
Since electrons are negatively charged, this gives
\begin{eqnarray}
X_i = \exp[\alpha_i V_i] 
\label{Eq:X_exp_V}
\end{eqnarray}
with $\alpha_i= -\big({{\rm d}\kappa \big/ {\rm d}V_i}\big) >0$,
where  $V_i=0$ is chosen to coincide with $X_i=1$.  
We mainly work with Eq.~(\ref{Eq:X_exp_V}),
but the almost topological fractional pumping also holds for $X_i=\exp[f_i(V_i)]$
for any $f_i(V_i)$ which is very positive for  $V_i \to \infty$, 
and is very negative for  $V_i \to -\infty$ (see the end of Sec.~\ref{sec:lowT-pump}).
This covers many physical systems.

\section{Adiabatic almost-topological pumping of a fraction of an electron per cycle}
\label{sec:describing-results}

%%%%%%%%%%%%%%%%%%%%%%%%%%%%
\begin{figure}
\includegraphics[width=0.3\textwidth]{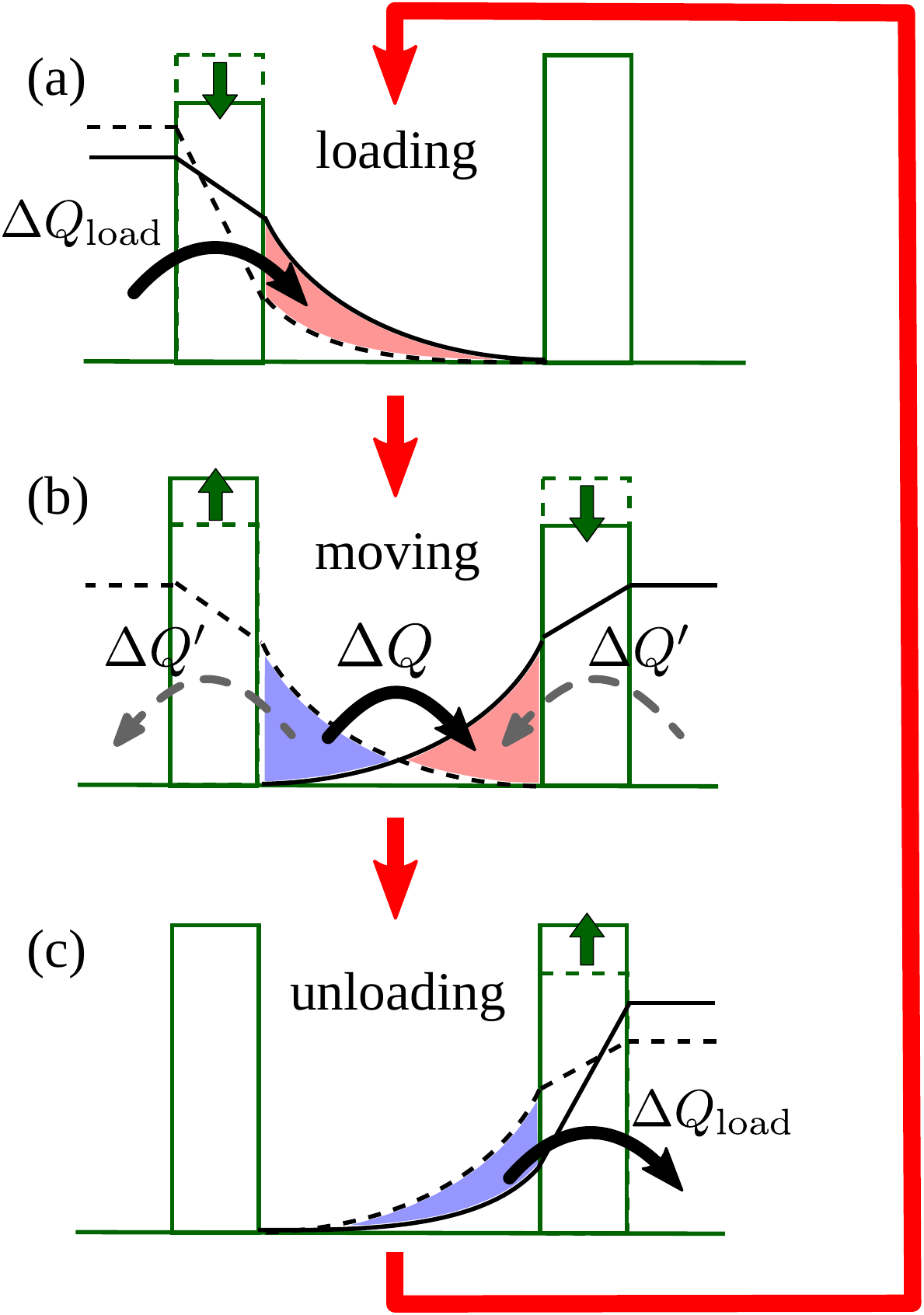}
\caption{\label{Fig:physical-wide-band}
Cartoon of the three steps corresponding to 1a, 1b and 1c , described in 
section~\ref{sec:wideband-explanation}, for a system without a Lamb shift ($\lambda=0$). 
The central region represents the quantum dot, 
separated from the reservoirs by barriers, whose height we can vary to change the tunnel-coupling between the dot and reservoirs.  Although the dot is a single site, it helps our intuition to show the dot's hybridization with reservoir L (R) as a decaying wavefunction penetrating the dot from the left (right).
Pink represents the average occupation of the state increasing with time, 
while blue represents it decreasing. The arrows indicate the average charge flow;
the arrows in (b) indicate the charge  $\Delta Q_{\rm load}$ is split in two,
with $\Delta Q'= \Delta Q_{\rm load}-\Delta Q$ going back into L, being replaced by a charge $\Delta Q'$ from R, see section \ref{sec:dot-occupation}.
}
\end{figure}
%%%%%%%%%%%%%%%%%%%%%%%%%%

Let us now briefly overview our main results, with the detailed calculations postponed to 
Sec.~ \ref{sec:general-results}.
Firstly, for a dot coupled to reservoirs without band-structure, 
there is a topological pumping at half an electron per cycle.
Secondly, one can choose the reservoir band-structure to ensure the pumped charge 
is topologically quantized at an arbitrary fraction of an electron per cycle.

\subsection{Half an electron per cycle}
\label{sec:wideband-explanation}
Here we consider a situation where the reservoir density of states is energy independent ($\omega$-independent), which is know as the wide band limit, and so $\rho(\omega)=\rho$.  
Then the reservoir induces a level-broadening of the quantum dot's energy level, 
but induces no Lamb shift; $\Lambda(\omega;\bm{K})=0$ in Eq.~(\ref{Eq:Lambda}).
Our calculations (using scattering theory in section \ref{sec:scattering} or Keldysh theory in section~\ref{sec:Keldysh}) show that this control of the level-broadening allows the pumping 
of half an electron per cycle in the low temperature limit.  

The dot level is taken to be above the reservoir's electro-chemical potential,
$(\epsilon_{\rm d} - \mu) > 0$,
and the pumping cycle is taken to be cycle 1 in Fig.~\ref{Fig:system}b,c, 
with neither $\epsilon_{\rm d}$ nor $\mu$ change during the pumping cycle.  
The basic physical process, sketched in Fig.~\ref{Fig:physical-wide-band} is the following:
\begin{itemize}
\item[(a)]  
Loading (segment 1a in Fig.~\ref{Fig:system}c): 
The dot starts weakly coupled to the reservoirs ($V_{\rm L}$ and $V_{\rm R}$ very negative) so the dot's level-broadening is much less than 
$(\epsilon_{\rm d} - \mu)$, as a result the dot's occupation is negligible.
The coupling to reservoir L is increased  ($V_{\rm L}$ increased), so that 
the reservoir wavefunctions spread into the dot (as in Fig.~\ref{Fig:physical-wide-band}a), as the dot state hybridizes with reservoir states.  The dot thus absorbs a charge of $\Delta Q_{\rm load}$.  Once the level-broadening is much more than $(\epsilon_{\rm d} - \mu)$, one reaches the limit where half the broadened level is below the reservoir's Fermi energy.
In this limit, there is half an electron in the dot, $\Delta Q_{\rm load}\to\half$; in other words a 50\% chance of finding the dot level occupied.
\item[(b)] 
Moving  (segment 1b in Fig.~\ref{Fig:system}c):
The coupling to reservoir L is slowly reduced to zero, 
while that to reservoir R is slowly increased to its maximum value  ($V_{\rm L}$ reduced and $V_{\rm R}$ increased), 
in such a way that the sum of the two couplings remains constant. 
Thus, the wavefunctions of reservoir R spread more into the 
dot, while those of reservoir L spread less into the dot.  
The occupation of the dot remains the same, but the dot state's hybridization 
moves from reservoir L to reservoir R. 
\item[(c)]  
Unloading  (segment 1c in Fig.~\ref{Fig:system}c):
The coupling to R is reduced ($V_{\rm R}$ reduced) so the level-broadening again becomes much less than $(\epsilon_{\rm d} - \mu)$.
As a result, the dot level empties into reservoir R, as the reservoir wavefunctions spread into the dot become negligible, and one returns
the dot to its initial state.
\end{itemize}
This cycle transfers a charge of $\Delta Q$ from reservoir L to reservoir R,
with $\Delta Q \neq  \Delta Q_{\rm load}$.
When the coupling is large enough that the level-broadening in step 1b is much more than $(\epsilon_{\rm d} - \mu)$, then $\Delta Q \to \Delta Q_{\rm load} \to 1/2$.

%==============================================
\begin{figure}
\begin{center}
\includegraphics[width=0.95\columnwidth]{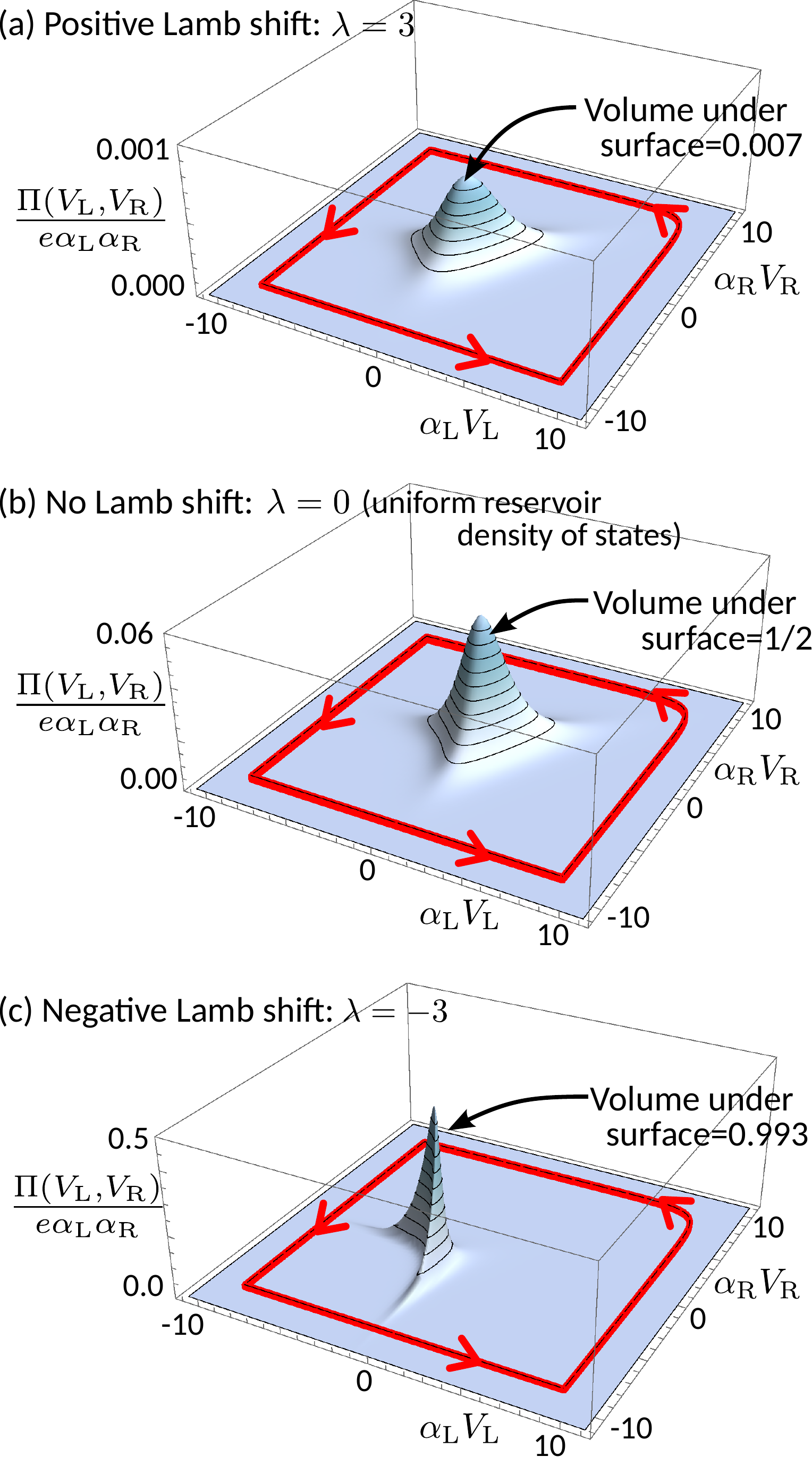}
\caption{\label{fig:curvature-versus-Vgate}
Plots of the Berry curvature, $\Pi_{\rm R}(V_{\rm L},V_{\rm R})$, for the dot-reservoir coupling in Eq.~(\ref{Eq:X_exp_V}).
This is given by Eq.~(\ref{Eq:Pi-versus-Vs-wideband}) for $\lambda=0$ and by 
Eq.~(\ref{Eq:Pi-versus-Vs}) for arbitrary $\lambda$.
It is always a sharp peak,
but the volume under the peak is highly $\lambda$-dependent, and given by Eq.~(\ref{Eq:triangle-contour-limit}).  Contour 1 from Fig.~\ref{Fig:system}b is also shown. 
}
\end{center}
\end{figure}
%==============================================

\subsection{Seeing the topology}
\label{sec:why-quantized}

The adiabatic charge pumped per cycle can be said to be topological when it is the same for all adiabatic pumping cycles of  the gate voltages that have the same topology.  
We will show this is the case for the cycle of $V_{\rm L}$ and $V_{\rm R}$  outlined above
under certain conditions, and up to exponentially small corrections; so we call it ``adiabatic almost-topological''  pumping.

To see what this means, one must write the charge pumped into R
as the surface integral, 
\begin{eqnarray}
\Delta Q_{\rm R} = e \int_{I(C)} {\rm d}V_{\rm L} {\rm d}V_{\rm R} \ \Pi_{\rm R}\left[V_{\rm L},V_{\rm R}\right]
\label{Eq:DeltaQ-integral-over-V}
\end{eqnarray}
where $I(C)$ is the surface in the $V_{\rm L}$-$V_{\rm R}$ plane enclosed by the pumping cycle $C$,

Then one calculates $\Pi_{\rm R}\left[V_{\rm L},V_{\rm R}\right]$, which is known as the Berry curvature, for the pumping. 
If one finds that this Berry curvature  
is a Dirac $\delta$-function, then the pumping is entirely topological; the adiabatically pumped charge only depends on how many times the  pumping contour winds around the $\delta$-function.
Here, our central result, Eq.~(\ref{Eq:Pi-versus-Vs}), is that the Berry curvature is not a $\delta$-function, but it is strongly peaked  with an exponential decay away from the peak, see Fig.~\ref{fig:curvature-versus-Vgate}.  Then we call the pumping {\it almost topological}, because it depends only on the contour's topology (how many times it winds around the peak) if the contour stays away from the peak, and if we neglect the exponentially small corrections coming from the tail of the peak.
Thus contours 1 and 2 in Fig.~\ref{Fig:system}b pump the same charge
(up to exponentially small corrections) because they both have the same topology --- each winds once around the peak. 

Fig.~\ref{fig:curvature-versus-Vgate}b shows the peak for reservoirs with uniform density of states. The integral over this peak is $1/2$, so the contours in Fig.~\ref{fig:curvature-versus-Vgate}b will thus pump the charge  
\begin{eqnarray}
\Delta Q_{\rm quantized} = e/2.
\end{eqnarray}

In the limit of thick tunnel barriers, $L\to \infty$, one sees that $\alpha_i$ in Eq.~(\ref{Eq:X_exp_V}) also goes to infinity.
Then the Berry curvature peak becomes a Dirac $\delta$-function in the  $V_{\rm L}$-$V_{\rm R}$ plane.
This means that the adiabatic pumping will becomes {\it entirely topological}.
However, for $L\to \infty$, the tunnel coupling is exponentially small,
so we require exponentially small temperatures, so $\eps_{\rm D}-\mu$ can be as small as the couplings, to ensure we can make 
$X_{\rm L}$ and $X_{\rm R}$ of order one, so the pumping contour can enclose the $\delta$-function peak.

\subsection{Different fractions of an electron per cycle}

Let us now consider reservoirs with a non-uniform density of states, so $\rho(\omega)$ depends on $\omega$.  In this case, the Lamb shift in Eq.~(\ref{Eq:Lambda}) is non-zero; this means that the dot-reservoir coupling does not only broaden the dot-level into a resonance, it also causes the centre of that resonance to be shifted in energy.
Sec.~\ref{sec:general-results} will use Keldysh theory to show that the adiabatic almost-topological pumping is quantized at a fraction of an electron (between 0 and 1), which is given by the ratio of the Lamb shift to the level-broadening.
We define $\lambda$ as the following dimensionless measure of this ratio
at $\omega=\mu$,
\begin{eqnarray}
\lambda= {2\Lambda(\mu,t) \big/\Gamma(\mu,t)},
\label{Eq:lambda}
\end{eqnarray}
where the factor of 2 is to make our results compact.
We will show that the almost topological charge that is pumped by the cycle described in Sec.~\ref{sec:wideband-explanation} above is
\begin{eqnarray}
\Delta Q_{\rm quantized} &=&  {e \over \pi}\left[ {\pi \over 2} -\arctan(\lambda)-  {\lambda \over 1+\lambda^2}  \right] . \quad
\label{Eq:triangle-contour-limit}
\end{eqnarray}
Hence, for this pumping cycle, $\Delta Q_{\rm quantized}$  is a monotonically decaying function of $\lambda$, and it take values between $e$ and 0. 
More precisely, $\Delta Q_{\rm quantized}$ equals $\left[1-2\big/(3\pi\lambda^2)\right]e$ for $\lambda\ll -1$, equals $e/2$ at $\lambda=0$, and 
equals
$2e\big/(3\pi\lambda^2)$ for $\lambda\gg 1$.

It is surprising that the exact result for pumping at low-temperatures only depends on the ratio of the Lamb shift
{\it at the electro-chemical potential} to the level-broadening {\it at the electro-chemical potential}, when many other observables depend on these quantities integrated over all energies (see e.g.\ $n(\bm{K})$ is Sec.~\ref{sec:dot-occupation}).  It is not easy to explain how this quantity emerges in the exact calculation, but we believe it is because 
we are at very low temperature and zero bias, so all charge flow between reservoirs occurs at energies at (or extremely close to) the electro-chemical potential.  Hence the pumped charge also only depends on the physics of the  Lamb shift
and level-broadening at the electro-chemical potential.

Crucially, $\lambda$ is entirely determined by the reservoir band-structure
since Eqs.~(\ref{Eq:Gcoupling}), (\ref{Eq:Lambda}) and (\ref{Eq:lambda}) mean that
\begin{eqnarray}
\lambda = {2 \over \rho(\mu)}\ \,P\hskip -3.6mm \int {\rm d} \varepsilon \,{ \rho (\varepsilon) \over \mu - \varepsilon} ,
\label{Eq:lambda-as-integral}
\end{eqnarray}
so it is independent of $K_{\rm L}$, $K_{\rm R}$ and $t$.
Hence, any given reservoir band-structure will have a given $\lambda$, and hence a given
quantized fraction of an electron pumped per cycle. 
By choosing a suitable reservoir band-structure,
one can choose that fraction.

\subsection{Requirements for experimental observation}
\label{sec:requirements}

There are four requirements for observing this quantized pumping of a fraction of an electron per cycle.

The first requirement is a quantum dot that mimics the Hamiltonian in Eq.~(\ref{Eq:H}),
which neglects electron-electron interactions on the dot. The simplest experimental implementation of Eq.~(\ref{Eq:H}) is an interacting quantum dot (described by an Anderson impurity Hamiltonian) in a large enough magnetic field that the dot's spin-state with higher energy is always empty, which makes the on-dot interaction term negligible.

The second requirement is that $k_{\rm B}T$ is much smaller than $(\epsilon_{\rm d}-\mu)$, 
larger temperatures will destroy the quantization. At the same time $(\epsilon_{\rm d}-\mu)$ should be small enough that we can make the dot-reservoir coupling $K \gg (\epsilon_{\rm d}-\mu)/\rho(\mu)$.
Thus we require that $k_{\rm B}T \ll K_{\rm max}\rho(\mu)$, which means the required value of $T$ depends on how strongly the dot can be coupled to the reservoirs.

The third requirement is related to the fact that the charge pumping is probabilistic,
with only the {\it average} charge being quantized.  This probabilistic nature of the pumping
is typical whenever there is part of the pumping cycle in which the dot is coupled to both reservoirs at the same time (segment 1b of the cycle). 
Thus in any given cycle $n=0,\pm 1,\pm2, \cdots$ electrons might flow.   
Central limit theorem tells us that averaging over many cycles will give an answer that will  converge to the quantized fraction that we predict.

The fourth requirement is due to our assumption that $\epsilon_{\rm d}$ is time-independent during the pumping cycle. Unfortunately, in practice, the electrostatic gates that vary  $K_{\rm L}$ and $K_{\rm R}$,
will also have a capacitive coupling to the dot-level,
causing $\epsilon_{\rm d}$ to vary.
Gate M in Fig.~\ref{Fig:system}a will minimize this capacitive coupling, by
partially screening the dot from gates L and R.  
Any remaining capacitive coupling to gates L and R will act much like the Lamb shift.
However, this coupling goes linearly in $V_{\rm L}$ and $V_{\rm R}$, while  the level-broadening and Lamb shift (if present) go exponentially, as above Eq.~(\ref{Eq:X_exp_V}).  Hence any effect of the capacitive coupling on $\eps_{\rm d}$ will become negligible compared to the broadening at large $\alpha_i V_i$.  

\section{Scattering theory}
\label{sec:scattering}

The central calculation in this work uses the Keldysh technique,
however as a warm up exercise, we can use the scattering theory of quantum pumping\cite{Brouwer1998Oct}
for the special case where the reservoirs have uniform density of states.
Readers interested in the Keldysh calculation of the general
case can skip this section.

The scattering matrix of a single-level dot (see e.g.~Refs.~\onlinecite{Fyodorov1997Apr,Alhassid2000Oct}) at energy $\mu$ is 
\begin{eqnarray}
\left(\!\begin{array}{cc} {S}_{\rm LL} & {S}_{\rm RL} \\ {S}_{\rm LR} & {S}_{\rm RR} \end{array}\!\right) \,=\, 
\bm{I}
- {i \over \mu-\epsilon_{\rm d} +i\half \Gamma} 
\left( \!\begin{array}{cc} 
\Gamma_{\rm L} & \sqrt{\Gamma_{\rm L}\Gamma_{\rm R}} \\
 \sqrt{\Gamma_{\rm L}\Gamma_{\rm R}} &  \Gamma_{\rm R}
\end{array}\!\right), \quad
\nonumber \\
\label{Eq:S-single-level-dot}
\end{eqnarray}
where $\bm{I}$ is a 2-by-2 unit matrix.
The scattering theory\cite{Brouwer1998Oct} for pumping of $K_{\rm L}$ and $K_{\rm R}$
around the contour $C$, 
gives the charge pumped per cycle into reservoir R as the integral over the surface enclosed by $C$, 
\begin{eqnarray}
\Delta Q_{\rm R} = e \int_{\rm C} {\rm d} K_{\rm L}{\rm d} K_{\rm R} \ \Pi_{\rm R}(\bm{K})
\label{Eq:scattering1}
\end{eqnarray}
where the Berry curvature $\Pi_{\rm R}(\bm{K})$ for our system is
\begin{eqnarray}
\Pi_{\rm R}(\bm{K}) &=& {1\over \pi}\,{\rm Im}\left[ 
{\partial {S}^*_{\rm RL}\over \partial K_{\rm L}} \,{\partial  {S}_{\rm RL}\over \partial K_{\rm R}}
+
{\partial {S}^*_{\rm RR}\over \partial K_{\rm L}} \,{\partial {S}_{\rm RR}\over \partial K_{\rm R}} \right].
\label{Eq:scattering2}
\end{eqnarray}
Substituting in Eq.~(\ref{Eq:S-single-level-dot}) and using Eq.~(\ref{Eq:rescale}),
one find that  the zero-temperature result for pumped charge per cycle (in units of $e$) is given by the dimensionless integral
\begin{eqnarray}
{\Delta Q_{\rm R} \over e} =  {2\over \pi} \int_{I(C')} d X_{\rm L} dX_{\rm R}  \,{X \over  \left[1+X^2\right]^2},
\label{Eq:analytic-result-any-C'-wb}
\end{eqnarray}
where $X=X_{\rm L}+X_{\rm R}$. The surface of integration $I(C')$ is that enclosed by the contour $C$ in Eq.~(\ref{Eq:scattering1}) rescaled using Eq.~(\ref{Eq:rescale}).  One can show\cite{Brouwer1998Oct} that $\Delta Q_{\rm L}= -\Delta Q_{\rm R}$.

Now we cast this result in terms of the gate-voltages that 
control the couplings.  Using Eq.~(\ref{Eq:X_exp_V}), $\Delta Q_{\rm R}$ is given by Eq.~(\ref{Eq:DeltaQ-integral-over-V})
with the Berry curvature
\begin{eqnarray}
{ \Pi_{\rm R}[V_{\rm L},V_{\rm R}] \over e}=  {2\alpha_{\rm L}\alpha_{\rm R}\over \pi}\, 
{\e^{\alpha_{\rm L}V_{\rm L}}\,\e^{\alpha_{\rm R}V_{\rm R}} \left(\e^{\alpha_{\rm L}V_{\rm L}}+\e^{\alpha_{\rm R}V_{\rm R}}\right)
\over  \big[1+\left(\e^{\alpha_{\rm L}V_{\rm L}}+\e^{\alpha_{\rm R}V_{\rm R}}\right)^2\big]^2}. \qquad
\label{Eq:Pi-versus-Vs-wideband}
\end{eqnarray}
This is shown in Fig.~\ref{fig:curvature-versus-Vgate}b.
The crucial feature  is that this is highly peaked at small $|\alpha_iV_i|$ and decays exponentially with increasing $|\alpha_iV_i|$
(for both $i={\rm L}$ and $i={\rm R}$).  Hence it fulfils the conditions for adiabatic almost-topological  pumping 
discussed in section~\ref{sec:why-quantized}.  

To find the charge pumped by a contour that encloses the above peak once,
we take contour 1 in Fig.~\ref{Fig:system}b, whose segment 1b is chosen such that 
$\exp[\alpha_{\rm L}V_{\rm L}]+\exp[\alpha_{\rm R}V_{\rm R}]$ is constant.
We then go back to Eq.~(\ref{Eq:analytic-result-any-C'-wb}), for which this contour maps  via Eq.~(\ref{Eq:X_exp_V}) to 
the triangular contour shown in Fig.~\ref{Fig:system}c.
The contour $C'$ is the triangle defined by $(X_{\rm L},X_{\rm R})$ going from $(0,0) \to (X_{\rm max},0) \to (0,X_{\rm max}) \to (0.0)$, where
$X_{\rm max} = \rho K_{\rm max}\big/\big[2(\epsilon_{\rm d}-\mu)\big]$.
We write 
\begin{eqnarray}
\int_{I(C')} d X_{\rm L} d X_{\rm R} \,(\cdots) \ =\  \half \int_0^{X_{\rm max}}d X \int_{-X}^{X} dY \, (\cdots), \qquad 
\label{Eq:transform_to_XY}
\end{eqnarray}
where $Y=X_{\rm L}-X_{\rm R}$.  Then
\begin{eqnarray}
{\Delta Q_{\rm R} \over e}
%&=& {2 \over \pi}\int_0^{X_{\rm max}} { dX \ X^2 \over  \left[1+X^2\right]^2}
%\nonumber \\
&=& {1 \over \pi} \left[ \arctan[X_{\rm max}] - {X_{\rm max} \over 1+X_{\rm max}^2}\right] 
\label{Eq:analytic-result-wb}
\end{eqnarray}
for the above triangular contour.
This goes to $1/2$ for large $X_{\rm max}$, 
which corresponds to encircling the peak in Eq.~(\ref{Eq:Pi-versus-Vs-wideband}).
Hence,  for uniform reservoir
density of states, the pumping is quantized at half an electron per cycle.

We do not know of a scattering theory for non-uniform reservoir density of states,
so we use the Keldysh formalism to treat such cases in sections~\ref{sec:Keldysh}-\ref{sec:general-results}.

\section{Comparison with dot occupation}
\label{sec:dot-occupation}

One might naively guess that the pump is simply due to filling the dot state 
from L in
the ``loading'' part of the cycle,
and then emptying it into R in the ``unloading'' part of the cycle. 
Then the charge transferred from L to R would 
equal the charge loaded into the dot, $\Delta Q_{\rm load}$. 
We show here that this is not the case; there is no simple relation between the pumped charge and $\Delta Q_{\rm load}$.

We are pumping adiabatically slowly, so electrons are continuously tunnelling in and out of the dot from L and R (and tunnelling though the dot from L to R) during the ``moving'' part of the cycle.  
They have too little energy to remain in the dot, 
but the uncertainty principle means they can be there for a time of order $\hbar\big/(\epsilon_{\rm d} - \mu)$.  
So there is no reason to assume the pumped charge is related to the dot occupation.
Indeed, the occupation of the dot at low temperatures, see e.g. Refs.~\onlinecite{Yang2015Oct,Jussiau2018Nov},
is
\begin{align}
	n(\bm{K}) &= \int_{-\infty}^\mu \frac{d\omega}{2\pi} \frac{\Gamma (\omega;\bm{K})}
	{\big[\omega- \epsilon_{\rm d} - \Lambda(\omega;\bm{K})\big]^2 +\left[\half\Gamma(\omega;\bm{K})\right]^2}.
	\nonumber
\end{align}
For a uniform density of states $\Lambda(\omega)=0$, the integrand is a Lorentzian, and so
$n(\bm{K}) =\arctan\left[X\right]\big/\pi$.  
Then 
\begin{eqnarray}
\Delta Q_{\rm load}\,=\,e\big[n(\bm{K}_{\rm max}) -n(0)\big]\,=\,{e\,\arctan\left[X_{\rm max}\right]\over \pi}.\qquad
\end{eqnarray}
From Eq.~(\ref{Eq:analytic-result-wb}), 
we see the pumped charge is smaller than $\Delta Q_{\rm load}$
by a factor of $\Delta Q'= e X_{\rm max}\big/\big[\pi(1+X_{\rm max}^2)\big]$,
which vanishes when $X_{\rm max}\to\infty$.
This means that the ``moving'' part of the pumping cycle  in section \ref{sec:wideband-explanation} involves a small flow, $\Delta Q'$, from the R to L through the dot (the dashed arrows in the Fig.~\ref{Fig:physical-wide-band}b).

For non-uniform density of states, $\Delta Q_{\rm load}$ depends on the $\omega$-dependence of $\Gamma (\omega;\bm{K})$ and $\Lambda (\omega;\bm{K})$ for all $\omega\leq \mu$.
In contrast,  the pumped charge in Eq.~(\ref{Eq:triangle-contour-limit}) 
depends {\it only} on their values at $\omega=\mu$.
Thus in general $\Delta Q$ and $\Delta Q_{\rm load}$ will not be related in any way, although both will be between $0$ and $e$.
Either can be larger, so $\Delta Q'$ can be of either sign. 
Indeed, two different set-ups can have the same $\Delta Q$ and different $\Delta Q_{\rm load}$, or vice-versa.

\section{Keldysh formalism}
\label{sec:Keldysh}

%=================
\begin{figure}
\begin{center}
\includegraphics[width=0.8\columnwidth]{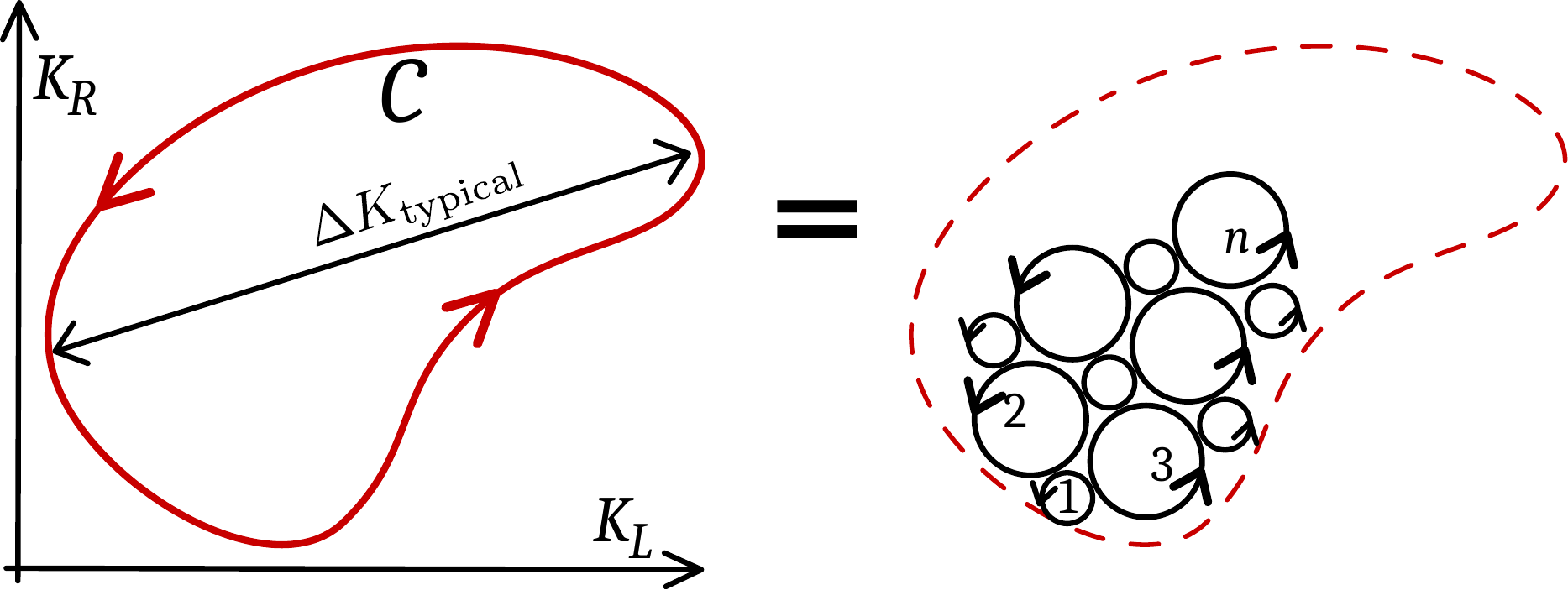}
\caption{\label{fig:contour_divide}
A sketch of the subdivision of contour $C$ into infinitesimal circular contours $\{C_n\}$, with the diameter of each infinitesimal contour chosen to ensure $C$ is filled densely.
}
\end{center}
\end{figure}
%=================

The dot's occupation and current into reservoir $i$ at time $t$ 
are\cite{Jauho94,Haug2008,Fransson2010,Hasegawa2017Jan,Hasegawa2018Feb}
\begin{align}
	n(t)& = \Braket{d^{\dagger}(t) d(t)} = - i G^<(t,t) ,\\
	J_i(t)& = -e \frac{d}{dt} \sum_k \Braket{c^{\dagger}_{ik}(t)c_{ik}(t)} \\
	&= e \int dt_1 \Bigl[ G^{\rm R}(t,t_1) \Sigma_i^<(t_1,t) + G^<(t,t_1)\Sigma_i^A(t_1,t) \nonumber \\
	& \hspace{1.5cm} - \Sigma_i^{\rm R}(t,t_1) G^<(t_1,t) -  \Sigma_i^<(t,t_1) G^A(t_1,t) \Bigr] .
	\nonumber
\end{align}
respectively, in terms of the  Keldysh Green's functions in Appendix~\ref{sec:app-Keldysh}.
We will derive the pumped charge for a large driving contour $C$, by summing the contributions 
from all infinitesimal circular contours inside it $\{C_n\}$ 
(see Fig. \ref{fig:contour_divide}),  
as was done in scattering theory by Ref.~[\onlinecite{Brouwer1998Oct}].

The infinitesimal contour $C_n$ satisfies
\begin{align}
	\bm{K}(t) \equiv   \left(\begin{array}{c} K_{\rm L}(t) \\ K_{\rm R}(t) \end{array}\right)
	\ = \ \bm{K}_{n,0} + \delta K  \left(\begin{array}{c}\cos [\Omega t] \\ \sin [\Omega t]  \end{array}\right) , \label{eqn:small_driving_def}
\end{align}
where $\Omega$ is a pumping frequency, $\delta K $ is an infinitesimally small amplitude of driving around the time-independent point $\bm{K}_{n,0}$.
Under this infinitesimal driving, the time-dependent charge current into reservoir $i$ is 
\begin{align}
	J_i(t) &= \mathcal{G}_{i}^{\rm L}(\Omega;\bm{K}_{n,0}) \,\delta K \cos (\Omega t ) \nonumber \\
	&\hspace{1.0cm} + \mathcal{G}_{i}^{\rm R}(\Omega;\bm{K}_{n,0}) \,\delta K \sin (\Omega t) . \label{eqn:charge_current_with_dynamic_G}
\end{align}
where $ \mathcal{G}_{i}^{\rm L}(\Omega;\bm{K}_{n,0})$ is the Fourier transform of the dynamic conductance for the infinitesimal contour $C_n$;
\begin{align}
	\mathcal{G}_{i}^{j}(t,t_1;\bm{K}_{n,0}) = \left. \frac{\delta J_i(t)}{\delta K_{j}(t_1)} \right|_{\delta K = 0}
	\label{Eq:dyn-cond}
\end{align}
This is given in terms of Keldysh Green's functions in Appendix \ref{sec:app-Keldysh},
and it only depends on the time differences $(t-t_1)$ because it is evaluated for $\delta K=0$. 
We assume the condition for adiabatic driving; 
\begin{align}
	\Omega \ll \tau^{-1}  \label{eqn:adiabatic_condition}.
\end{align}
where  $\tau$ is the typical time for electrons in the dot to relax.
Then it is sufficient to take the dynamic conductance at leading order in the pumping frequency $\Omega$;
$\mathcal{G}_{i}^{i'}(\Omega;\bm{K}) = A_i^{i'}(\bm{K})\  \Omega   \ +\ {\cal O}[\Omega^2]$.
Substituting this into Eq.~(\ref{eqn:charge_current_with_dynamic_G}), and 
integrating $\Omega t$ from $0$ to $2\pi$, 
we find the
charge pumped per cycle on the infinitesimal contour $C_n$ is 
$\delta Q_{i,n} = \int_{C_n} \left[ A_{i}^{\rm L}(\bm{K}) dK_{\rm L} + A_{i}^{\rm R}(\bm{K}) dK_{\rm R} \right]$.

Summing all infinitesimal contours inside the large contour $C$, gives 
charge pumped per cycle around $C$ as
\begin{eqnarray}
\Delta Q_i = \oint_{C} d\bm{K} \cdot \bm{A}_{i}(\bm{K})
\label{Eq:Berry-connect}
\end{eqnarray}
where we define the {\it Berry connection} as the vector
$\bm{A}_{i}(\bm{K}) = \left( A_{i}^{\rm L}(\bm{K}) \, ,\,A_{i}^{\rm R}(\bm{K}) \right)$.
Re-writing this in terms of a surface integral using Stokes theorem,
we get 
\begin{align}
\Delta Q_i = \int_{I(C)} \rmd \bm{S} \cdot \bm{\Pi}_i (\bm{K}),
\end{align}
where $\bm{\Pi}_i (\bm{K}) = \nabla_{\!\bm{K}} \times \bm{A}_i(\bm{K})$ is the {\it Berry curvature}.
This integral is over the surface  $I(C)$ which is enclosed by the pumping contour $C$.
As this surface is the $K_x$-$K_y$ plane, only the component of $\bm{\Pi}_i (\bm{K})$
perpendicular to this plane contributes; we call this component 
\begin{eqnarray}
\Pi_i (\bm{K})&=& {{\rm d} \over {\rm d}K_{\rm L}} [\bm{A}^{\rm R}_i(K)] \,-\, {{\rm d} \over {\rm d}K_{\rm R}} [\bm{A}_i^{\rm L}(K)]  \, .
\label{Eq:curvature_from_connection}
\end{eqnarray}
we will calculate this for our model in the next section.

We end this derivation with an adiabaticity condition for the large contour $C$.
Given Eq.~(\ref{eqn:adiabatic_condition}) for the infinitesimal circular contours, adiabaticity on $C$ requires 
\begin{align}
	\left|{d \bm{K} \big/ d t}\right|  \  \ll\   \Delta K_{\rm typical}\big/ \tau ,
	\label{Eq:adiab-cond}
\end{align}
where $\Delta K_{\rm typical}$ is the typical scale of the contour  (see Fig.~\ref{fig:contour_divide}), and $\tau$ is the relaxation time of the dot state.   The magnitude of $1/\tau$  is discussed in 
Sec.~\ref{sec:ad_condition_be}.

\section{Results for our model.}
\label{sec:general-results}

For the Hamiltonian in Eq.~(\ref{Eq:H}), we find that 
the Berry connection in Eq.~(\ref{Eq:Berry-connect}) contains two terms
\begin{eqnarray}
 \bm{A}_i (\bm{K})  =  \bm{A}^{\rm broad}_i (\bm{K}) + \bm{A}^{\rm shift}_i (\bm{K})  
\label{Eq:Q-connection}
\end{eqnarray}
because $\bm{A}_i (\bm{K})$ involves a derivative with respect to $\bm{K}$, 
and that derivative can act on the level-broadening (giving $\bm{A}^{\rm broad}$)  
or the  Lamb shift (giving $\bm{A}^{\rm shift}$).  
If there is no Lamb shift then $\bm{A}^{\rm shift}_i (\bm{K}) =0$, while if the Lamb shift is much greater than the level-broadening, then Eq.~(\ref{Eq:Q-connection}) is dominated by  
$\bm{A}^{\rm shift}_i (\bm{K}) $.
The Keldysh calculations outlined in Appendix~\ref{sec:app-Keldysh} give
\begin{align}
	\left[ \bm{A}^{\rm broad}_i (\bm{K})  \right]_j  
	&= \int \frac{d\omega}{2\pi} \ \Bigl[ \Bigl(\mathcal{B}^2 - \quarter \mathcal{A}^2 \Bigr) f \Lambda_{i}  \nonumber \\
	&\hspace{1cm} - \half \mathcal{A} \mathcal{B} f \Gcoupling_{i} - \delta_{i,j} \mathcal{B}^{\prime} f \Bigr] \frac{\partial \Gcoupling}{\partial K_{j}}, \\
	\left[ \bm{A}^{\rm shift}_i (\bm{K})  \right]_j  
	&= \int \frac{d\omega}{2\pi} \ \Bigl[2 \mathcal{B} \mathcal{A} f \Lambda_{i} + \quarter \mathcal{A}^2 (f^{\prime} \Gcoupling_i - f \Gcoupling_{i}^{\prime}) \nonumber \\
	&\hspace{1cm} + \mathcal{B}^2 (f \Gcoupling_i)^{\prime} - \delta_{i,j}(\mathcal{A} f)^{\prime} \Bigr] \frac{\partial \Lambda}{\partial K_{j}}. 
\end{align}
where $i$ and $j$ are L or R, and $f = [1+ e^{(\omega-\mu)/T} ]^{-1}$ is the Fermi function.
The primed denotes the partial derivative with respect to $\omega$.
The quantities $\Lambda_i$ and $\Gcoupling_i$ are given in Eqs.~(\ref{Eq:Gcoupling_i}-\ref{Eq:Lambda}),
while $\mathcal{A} = 2 \mathrm{Im} [ G^A(\omega)]$ and $\mathcal{B} = \mathrm{Re} [ G^A(\omega)]$,
with $G^A(\omega)$ given in Eq.~(\ref{Eq:GA}).

Turning to the Berry curvature in Eq.~(\ref{Eq:curvature_from_connection}), 
we see that it contains two derivatives (with respect to $K_j$),
because $ [\bm{A}_i(\bm K)]_j$ contained one.
Hence $\Pi_i (\bm{K})$ contains three terms;  
a ``broad-broad'' term due to both derivatives acting on the broadening, 
a ``shift-shift'' term due to both derivatives acting the Lamb shift, 
and a ``shift-broad'' term with one derivative on each of them.
The ``shift-shift'' term turns out to be zero, showing that
the Lamb shift alone is not enough to do pumping.  Intuitively, this can be understood as the Lamb shift only moving the dot level, which is not enough to do pumping.
Hence 
\begin{eqnarray}
	\Pi_{\rm R}(\bm{K}) &=& 
	\Pi_{\rm R}^{\mbox{{\scriptsize broad-broad}}}(\bm{K})
	+\Pi_{\rm R}^{\mbox{{\scriptsize shift-broad}}}(\bm{K}),
\end{eqnarray}
and $\Pi_{\rm L}(\bm{K})=-\Pi_{\rm R}(\bm{K})$, with
\begin{align}
	\Pi_{\rm R}^{\mbox{{\scriptsize broad-broad}}}(\bm{K})
	&=
	\int \frac{d\omega}{4\pi}  \ f^{\prime}  \mathcal{A} \mathcal{B} 
	\ { \Gamma^2(\omega,\bm{K}) \over K^2} \, ,
	\\
	\Pi_{\rm R}^{\mbox{{\scriptsize shift-broad}}}(\bm{K}) 
	&= \int \frac{d\omega}{4\pi} \ f^{\prime}  \mathcal{A}^2  \ 
	{ \Gamma(\omega,\bm{K}) \Lambda(\omega,\bm{K})\over K^2}\, . 
\end{align}
where we have used the fact that $\Lambda$ and $\Gcoupling$ are proportional to $K=K_{\rm L}+K_{\rm R}$. A
 bit more algebra gives 
 \begin{align}
	&\Pi_{\rm R}(\bm{K}) \nonumber \\
	&= \frac{e}{2} \int \frac{d\omega}{2\pi} \frac{ (\omega-\epsilon_{\rm d})\, \rho^2(\omega)\, \Gamma(\omega;\bm{K}) \,  \left(\partial f\big/\partial \omega\right) }
	{\left[ \big[\omega- \epsilon_{\rm d} - \Lambda(\omega;\bm{K})\big]^2 +\left [ \half\Gamma(\omega;\bm{K}) \right]^2 \right]^2}.
	\label{Eq:Berry-curvature-final}
\end{align}
This depends on the sum of the couplings, $K=(K_{\rm L}+K_{\rm R})$, but not the difference $(K_{\rm L}-K_{\rm R})$.

\subsection{Low temperature pumping}
\label{sec:lowT-pump}

In the limit of small temperature,  we can make the approximation $\left(\partial f\big/\partial \omega\right) = -\delta(\omega-\mu)$ in Eq.~(\ref{Eq:Berry-curvature-final}).
To justify this approximation, one needs the other terms in the integrand of  Eq.~(\ref{Eq:Berry-curvature-final}) to vary little over the window of $\omega$ given by $\mu\pm k_{\rm B}T$.
Then, the Berry curvature is
 \begin{align}
	&\Pi_{\rm R}(\bm{K}) 
	%\nonumber \\
	%&
	\,=\, \frac{e}{4\pi} \frac{ (\epsilon_{\rm d}-\mu)\, \rho^2(\mu)\,\Gamma(\mu;\bm{K}) }
	{\left[ \big[\mu- \epsilon_{\rm d} - \Lambda(\mu;\bm{K})\big]^2 +\left [ \half\Gamma(\mu;\bm{K}) \right]^2 \right]^2}.
	\label{Eq:Berry-curvature-T=0}
\end{align}
Writing this in terms of $\lambda$ in Eq.~(\ref{Eq:lambda}),
the low-temperature result for pumped charge per cycle (in units of $e$) is given by the dimensionless integral
\begin{eqnarray}
{\Delta Q_{\rm R} \over e} =  {2\over \pi} \int_{I(C')} d X_{\rm L} dX_{\rm R} \,{X \over  \left[(1+\lambda X)^2+X^2\right]^2}
\label{Eq:analytic-result-any-C'}
\end{eqnarray}
where $X_i$ is defined in Eq.~(\ref{Eq:rescale}), with $\rho$ being $\rho(\mu)$ and  $X=X_{\rm L}+X_{\rm R}$.
The surface of integration $I(C')$ is that enclosed by the contour $C$ in Eq.~(\ref{Eq:scattering1}) rescaled using Eq.~(\ref{Eq:rescale}).

%==============================================
\begin{figure}
\begin{center}
\includegraphics[width=0.95\columnwidth]{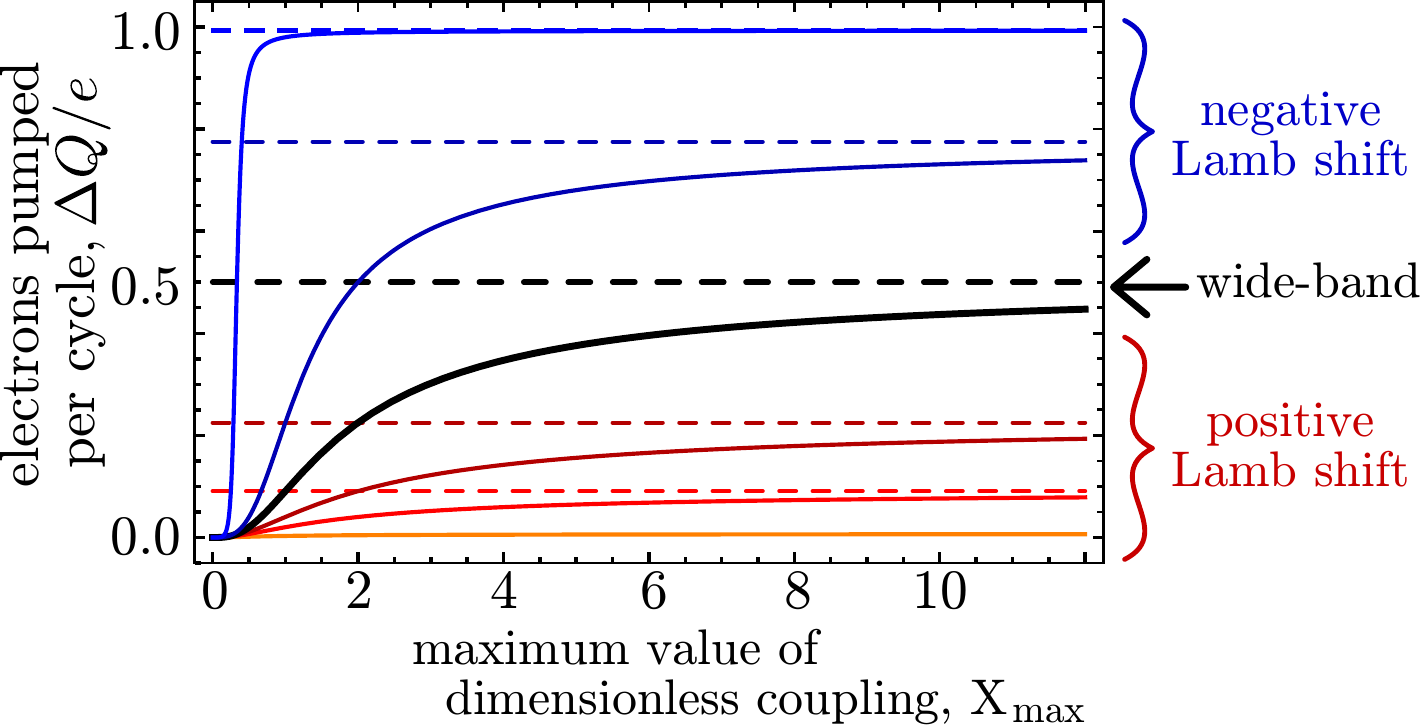}
\caption{\label{fig:general-result}
The solid curves are the charge pumped per cycle on the triangular pumping cycle, given by Eq.~(\ref{Eq:general-result-triangle-contour}).
From top to bottom we have $\lambda=-3,-0.5,0,0.5,1,3$.
The horizontal dashed lines show the large $X_{\rm max}$ limit given by Eq.~(\ref{Eq:triangle-contour-limit}).
}
\end{center}
\end{figure}
%==============================================

As explained in Sec.~\ref{sec:model}, we control gate-voltages $V_i$, in experiments. By substituting 
Eq.~(\ref{Eq:X_exp_V}) into Eq.~(\ref{Eq:analytic-result-any-C'}),
we find the Berry curvature in the $(V_{\rm L},V_{\rm R})$-plane 
\begin{eqnarray}
{\Pi_{\rm R}(V_{\rm L},V_{\rm R}) \over e}= {2\over \pi} \left.{\alpha_{\rm L}\alpha_{\rm R} \,X \, \e^{\alpha_{\rm L}V_{\rm L}}\e^{\alpha_{\rm R}V_{\rm R}} \over  \left[(1+\lambda X)^2+X^2\right]^2}\right|_{\hskip -3mm{X=\e^{\alpha_{\rm L}V_{\rm L}}\atop\hskip 7mm+\e^{\alpha_{\rm R}V_{\rm R}}}}, \qquad
\label{Eq:Pi-versus-Vs}
\end{eqnarray}
shown in Fig.~\ref{fig:curvature-versus-Vgate}.
This is our {\it central result}, because the fractional and topological nature of the adiabatic pumping both follow from it, as we now show.

Eq.~(\ref{Eq:Pi-versus-Vs}) has a peak at small $|\alpha_iV_i|$, and decays exponentially as $|V_i |$ grows.
Hence, any pumping contour that encloses the peak without encroaching on it will give the same pumped charge per cycle (up to exponentially small corrections),
ensuring quantized pumping.

To calculate the charge pumped by such a cycle, we return to Eq.~(\ref{Eq:analytic-result-any-C'}) and consider a triangular contour $C'$ explained above Eq.~(\ref{Eq:transform_to_XY}).
Eq.~(\ref{Eq:X_exp_V})  means that for large $X_{\rm max}$ 
this triangular contour corresponds to contour 1 in  Fig.~\ref{Fig:system}b, 
that encloses the peak in $\Pi_{\rm R}(V_{\rm L},V_{\rm R})$.
We transform to $X$ and $Y$ as in Eq.~(\ref{Eq:transform_to_XY}), then
\begin{eqnarray}
{\Delta Q_{\rm R} \over e} &=&  
{1\over \pi}\bigg[ {\pi \over 2} -\arctan\left({1+\lambda X_{\rm max} \over X_{\rm max}}\right)
\nonumber \\
& & \qquad-\,{X_{\rm max}(1+\lambda X_{\rm max}) \over 1+X_{\rm max}^2 +\lambda X_{\rm max} (2+\lambda X_{\rm max})}
\bigg],\qquad
\label{Eq:general-result-triangle-contour}
\end{eqnarray}
see Fig.~\ref{fig:general-result}. It reduces to Eq.~(\ref{Eq:analytic-result-wb}) for $\lambda=0$, since $\arctan(x)+\arctan(1/x)={\rm sgn}[x]\pi/2$.  We take $X_{\rm max}\to\infty$
to get the pumping for a contour that corresponds to one enclosing the peak of Eq.~(\ref{Eq:Pi-versus-Vs}); this gives
Eq.~(\ref{Eq:triangle-contour-limit}).

This analysis has given us our main results; the adiabatic pumping is almost topological, and pumps a fraction of an electron (between 0 and 1) given by the value of $\lambda$, which is determined purely by the reservoir's band-structure.

To generalise to a voltage dependence of the form below  Eq.~(\ref{Eq:X_exp_V}), we substitute it into Eq.~(\ref{Eq:analytic-result-any-C'}). Then Eq.~(\ref{Eq:Pi-versus-Vs}) changes, but it remains strongly peaked with exponentially 
small tails. This  ensures that there is still adiabatic almost-topological  pumping. 
Further more, the faction pumped per cycle is the same for any voltage dependence, 
since it was calculated directly from Eq.~(\ref{Eq:analytic-result-any-C'}).

\section{Adiabaticity and band gaps}
\label{sec:ad_condition_be} 

Up to now this work has only discussed pumping in the adiabatic limit.
However, from the argument in section \ref{sec:Keldysh}, it is clear that a large but finite pumping period $T_{\rm period}\sim \Delta K_{\rm typical}\big/(d \bm{K}/ d t) \gg \tau$, will induce a non-adiabatic correction of order $(\tau/T_{\rm period})$.  This non-adiabatic correction is much larger than that in many proposals for topological pumping, in which
the topology makes the non-adiabatic corrections exponentially small at large $T_{\rm period}$.
Thus to observe the topological pumping in our system it is crucial to estimate $\tau$, 
and then choose the pumping to be slow enough  (large $T_{\rm period}$) to make 
corrections of order $(\tau/T_{\rm period})$ negligible.

It is simple to estimate $\tau$ when the reservoirs have uniform density of states,
since there the dot state decays at the rate given by the level-broadening in section~\ref{sec:model}; 
$1/\tau =\Gamma = (K_{\rm L}+K_{\rm R})\rho$.
For systems with non-uniform density of states we can place a lower bound on  $1/\tau$ by saying that $1/\tau\gtrsim K\rho_{\rm min}$, where $\rho_{\rm min}$ is the minimal value of the density of states.

\begin{figure}[t]
\begin{center}
\includegraphics[width=0.3\textwidth]{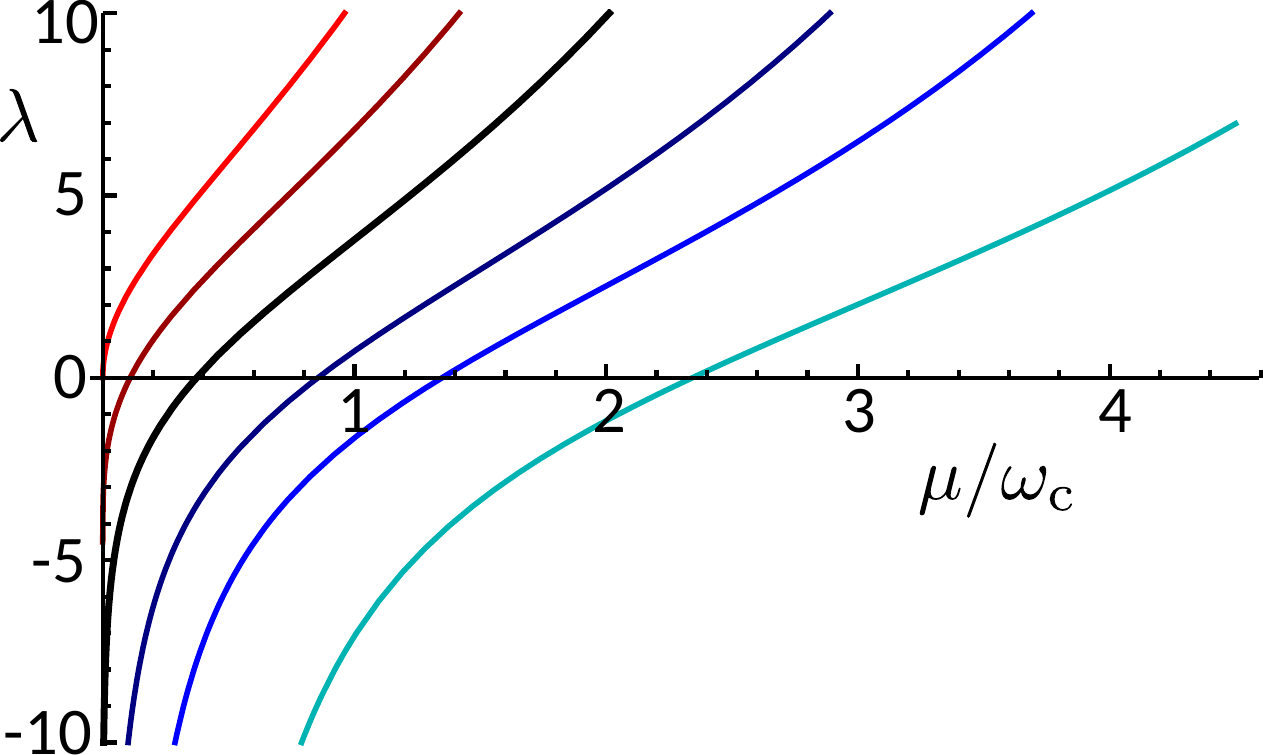}
\caption{\label{fig:lambda-for-bandgap}
The parameter $\lambda$ for a reservoir density of states given by Eq.~(\ref{Eq:rho-band-edge}) as a function of $\mu/\omega_{\rm c}$,
for $s=-0.5,-0.3,0,0.5,1$ and $2$ (from top to bottom).  One can access almost any $\lambda$ by a suitable choice of $s$ and $\mu/\omega_{\rm c}$.  
}
\end{center}
\end{figure}

However, this poses a problem for reservoirs with band-gaps, 
as the density of states vanishes in the band-gap, so the above lower-bound does not allow us
to say when the pumping is slow enough to be considered adiabatic.
To investigate this further we consider the case where the electro-chemical potential is near 
a band-edge in the reservoir, so the reservoir's density of states is  
\begin{align}
	\rho(\omega) =\left\{\begin{array}{ccl}
	\rho_0 \,\big(\omega\big/\omega_{\rm c} \big)^s \ e^{-\omega / \omega_{\rm c}} &\phantom{\Big|} & \omega >0
	\\
	0 &\phantom{\big|} & \omega <0
	\end{array}\right. ,
	\label{Eq:rho-band-edge}
	\end{align}
where, without loss of generality, we measure energy $\omega$ from the band-edge.
Then the level-broadening is $\Gcoupling_i=K_i \rho(\omega)$, and the Lamb shift is (see e.g.~Refs.~\onlinecite{Zhang2012,Jussiau2018Nov}), 
\begin{eqnarray}
\lambda= -2\,\Gamma(1+s)\ {\rm Re}\left[ (-1)^s \ \Gamma\left(-s,-\mu/{\omega_{\rm c}}\right)  \right].
\end{eqnarray}
Fig.~\ref{fig:lambda-for-bandgap} plots this, and shows that a suitable choice of $s$ and $\mu/\omega_{\rm c}$ will give almost any desired value of $\lambda$.

%=================
%=================

It has long been known that this model exhibits an infinite-lifetime bound-state\cite{Shiba1973Jul, John1990, John1991,John1994,Kofman1994}, see  Refs.~\onlinecite{Angelakis2004,Chang2018} for reviews.
Electron dynamics in various time-dependent versions of this model 
have been studied; particularly the decay of an 
initially prepared dot state\cite{Zhang2012,Xiong2015Aug,Tu2016Mar,Ali2015Dec,Lin2016,Ali2017Mar}, 
the response to switching on a bias,\cite{Dhar2006Feb,Stefanucci2007May}
or the response to periodic
driving.\cite{Jin2010Aug,Basko2017}
For $s >0$,  this bound-state appears when the coupling exceeds a critical value
\cite{Dhar2006Feb,Stefanucci2007May,Jin2010Aug,Xiong2015Aug,Yang2015Oct,Ali2015Dec,Tu2016Mar,Zhang2012,Engelhardt2016,Lin2016,Ali2017Mar,Jussiau2018Nov}
$K_{\rm c} =\epsilon/\Gamma[s]$.
This state has $\tau =\infty$, so pumping never satisfies the adiabaticity condition 
in Eq.~(\ref{Eq:adiab-cond}) when $K>K_{\rm c}$.
Intriguingly, the Berry curvature in  Eq.~(\ref{Eq:Berry-curvature-final}) does not contain $K_{\rm c}$; it is a smooth function across this line of critical coupling $K=(K_{\rm L}+K_{\rm R}) =K_{\rm c}$.
However, the Berry curvature in  Eq.~(\ref{Eq:Berry-curvature-final})  ceases to have a physical meaning when one crosses the line of critical coupling,
because non-adiabatic contributions dominate beyond this line ($K>K_{\rm c}$), 
no matter how slow the pumping is.

\begin{figure}
\includegraphics[width=0.36\textwidth]{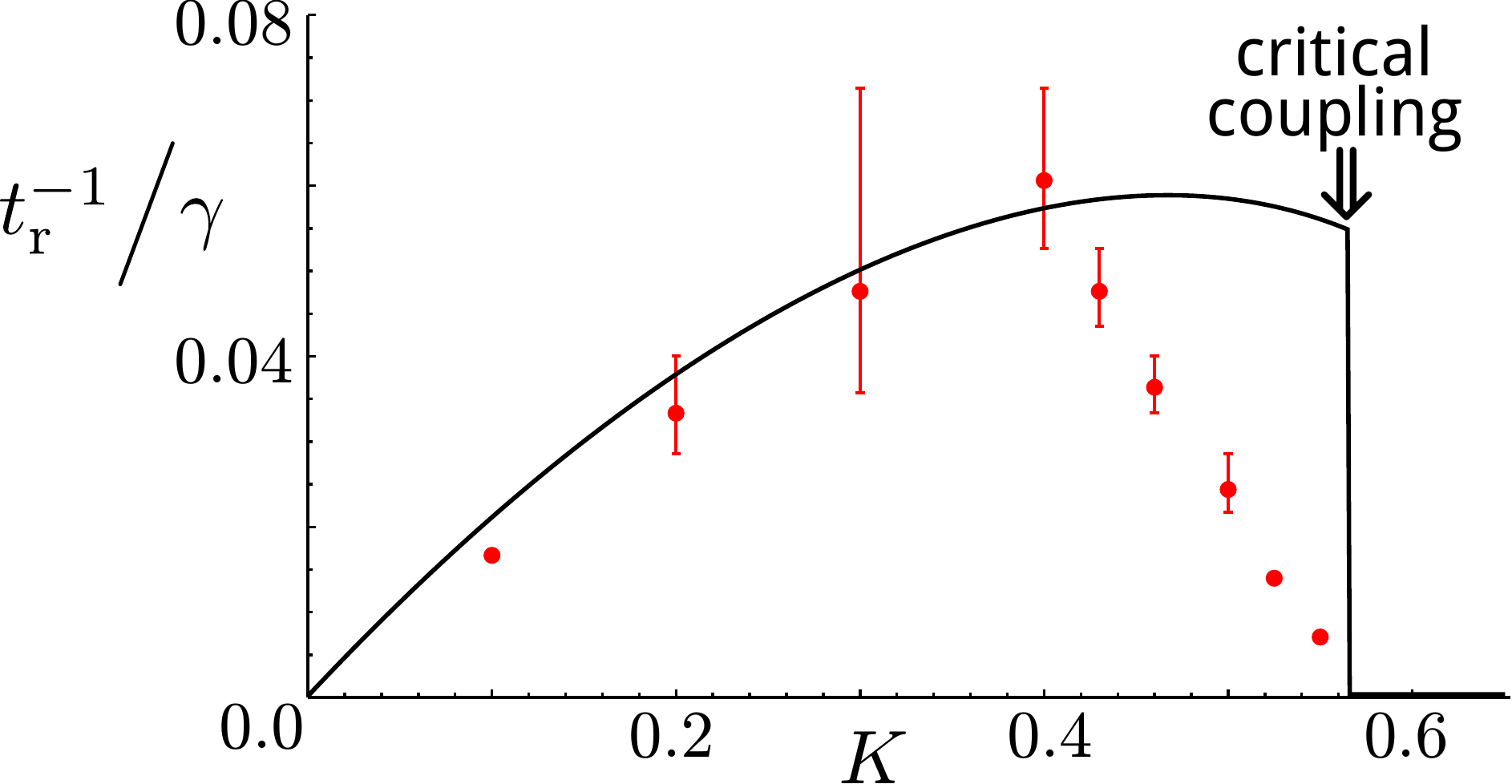}
\caption{\label{fig:RT002}
Estimate of the dot relaxation rate versus coupling $K$, for reservoirs with a band-gap; Eq.~(\ref{Eq:rho-band-edge}) with $s=1/2$ and $\omega_{\rm c}=10\eps_{\rm d}$.
The points estimate the rate via Eq.~(\ref{Eq:two-percent}).
The solid curve estimates it from the exponential part of the decay,
neglecting the oscillatory or powerlaw components.
At small $K$, the decay is almost exponential, and the two protocols coincide.  However, as $K \to K_{\rm c}$, the powerlaw completely dominates, and the solid curve  fails to capture the slowing of the decay. 
The infinite-lifetime bound-state emerges at $K_{\rm c}$; so $1/t_{\rm r}=0$ for $K\geq K_{\rm c}$.
Note the error-bars are not error-bars in the usual sense, see Sec.~\ref{sec:ad_condition_be}.
}
\end{figure}

For $K<K_{\rm c}$, it is difficult to determine the dot's decay rate, $1/\tau$,
because it contains terms with an oscillatory powerlaw decay, for which there is no unique way to
define $1/\tau$. Fig.~\ref{fig:RT002} is an attempt to give a feeling for how $1/\tau$
depends on the coupling.
The red data points are the inverse time for the dot occupation to decay to threshold 
(using the method reviewed in Refs.~\onlinecite{Yang2015Oct,Jussiau2018Nov}) that we set at 2\% of its initial value, i.e.
we plot the $1/t_{\rm r}$ that satisfies
\begin{eqnarray}
{|n(t_{\rm r})-n_0| \over |n_\infty-n_0|} = 2\% \, .
\label{Eq:two-percent}
\end{eqnarray}
The solid curve is the time taken to reach this threshold, if one approximates the decay to an exponential 
at the rate given by the imaginary part of the resonance's energy
(i.e. neglecting all powerlaw or oscillatory components in the decay).
This approximation captures much of the true decay, but misses the sharp drop in $1/t_{\rm r}$ as $K\to K_{\rm c}$.  This sharp drop shown by the data points indicates that the timescale to decay diverges as $K$ approaches $K_{\rm c}$.
Hence it is increasingly difficult to pump slowly enough to be adiabatic as $K$ gets closer to $K_{\rm c}$. 

The ``error bars'' on the data in Fig.~\ref{fig:RT002} are {\it not} numerical uncertainties
in $t_{\rm r}$ (such uncertainties are about the size of the red dots). 
They indicate the period of the oscillations in
the decay, which is maximal for $K \simeq K_{\rm c}/2$.
A small change in the system parameters (e.g. a change of $\omega_{\rm c}$ or $\mu$) would shift the phase of the oscillations, thereby shifting where the oscillating decay crosses the threshold
to a different place on the vertical ``error bar''.  Hence, we can expect a change in system parameters to induce a large change in $1/t_{\rm r}$ when $K \simeq K_{\rm c}/2$, while the change will be modest for $K\ll K_{\rm c}$ and $K\sim K_{\rm c}$.

%=================
%=================

\section{Conclusions}
\label{sec:conclusion}

We show that a system without exotic physics (a non-interacting single-level quantum dot at low temperature) can exhibit
an adiabatic  almost-topological pumping of a fraction of an electron per cycle, when averaged over many cycles.
We call it ``almost'' topological because the pumped charge depending only on the number of times the contour winds around the peak in the Berry curvature, shown in Fig.~\ref{fig:curvature-versus-Vgate},
under the conditions that (i) the contour does not touch the peak, and (ii) we neglect the exponentially small corrections coming from the tail of the peak.
Sec.~\ref{sec:why-quantized} mentions a specific limit in which the adiabatic pumping is entirely topological.
The fraction pumped (between zero and one electron) is determined by the
ratio of the Lamb shift to the level-broadening. This ratio is imposed by the reservoir band-structure,
which can be chosen to give the desired fraction. A uniform reservoir density of states gives the quantized pumping of half an electron per cycle. 
We emphasize that it is the {\it average} charge pumped per cycle that is (almost) topological and fractional.
Each cycle has a finite probability that $n$ electrons are pumped for $n=0,\pm1,\pm2,\cdots$; the quantized fraction is only revealed by averaging over many cycles.

Hence, if one wants to prove the existence of fractionally charged particles in some system,
one would need more evidence than just adiabatic pumping of fractional average charge.  
This evidence could be that non-adiabatic corrections decay exponentially when the period of the pumping cycle is made large, since these corrections only decay like one over this period in our model.

\section{Acknowledgements}
MH acknowledges the financial support of the Advanced Leading Graduate
Course for Photon Science. EJ and RW acknowledge the support of  the French National Research Agency program ANR-15-IDEX-02, via the Universit\'e Grenoble Alpes QuEnG project.

%{\bf Contributions:} MH derived all results using Keldysh, in discussion with EJ and RW. EJ reproduced the wideband results with scattering theory, and wrote the code for Fig.~\ref{fig:RT002}.  MH and RW wrote the manuscript.

%%%%%%%%%%%%%%%%%%%%%%%%%%%%%%
%%%%%%%%%%%%%%%%%%%%%%%%%%%%%%
\appendix

\section{Keldysh Green's functions}
\label{sec:app-Keldysh}

The quantum dot's Green's functions are defined as\cite{Jauho94,Haug2008,Fransson2010}
\begin{align}
	G^A(t_1,t_2) &= i \Theta (t_2-t_1) \Braket{[d(t_1),d^{\dagger}(t_2)]_+}, \\
	G^{\rm R}(t_1,t_2) &= -i \Theta (t_1-t_2) \Braket{[d(t_1),d^{\dagger}(t_2)]_+}, \\
	G^{<}(t_1,t_2) &= i \Braket{d^{\dagger}(t_2) d(t_1) } \label{eqn:def_les_GF} ,
\end{align}
where $\Theta (t)$ is a Heaviside function and $[\cdot,\cdot]_+$ is an anti-commutator.
Their algebraic form is given by Dyson's equations
\begin{align}
	G^\kappa(t_1,t_2) &= g^\kappa(t_1,t_2) \nonumber \\
	&\hspace{0.5cm} + \int dt_3 dt_4 \ g^\kappa(t_1,t_3) \Sigma^\kappa(t_3,t_4) G^\kappa(t_4,t_2) ,
	\nonumber
\end{align}
for $\kappa=R,A$, and 
\begin{align}
	G^{<}(t_1,t_2) &= \int dt_3 dt_4 \ G^R(t_1,t_3) \Sigma^<(t_3,t_4) G^A(t_4,t_2) .
	\nonumber
\end{align}
Here $g^{A/R}(t_1,t_2)$ are Green's function of electron of the isolated quantum dot, 
and $\Sigma^\kappa(t_1,t_2)$ are the one-particle-irreducible self-energy for $\kappa=A,R,<$, 
\begin{align}
	\Sigma_i^\kappa(t_1,t_2) &= \sum_k \gamma_{i}(t_1) g_{ik}^\kappa (t_1,t_2) \gamma_{i}(t_2) ,
\end{align}
with $\Sigma^\kappa=\Sigma^\kappa_{\rm L}+\Sigma^\kappa_{\rm R}$.
Here $g_{ik}^\kappa(t_1,t_2)$ is a Green's function of electrons in the isolated electron reservoirs, 
\begin{align}
	g_{ik}^A(t_1,t_2) &= i \Theta (t_2-t_1) \big\langle[c_{ik}(t_1),c_{ik}^{\dagger}(t_2)]_+\big\rangle_{\gamma_i=0} , \nonumber \\
	g_{ik}^{\rm R}(t_1,t_2) &= -i \Theta (t_1-t_2)  \big\langle [c_{ik}(t_1),c_{ik}^{\dagger}(t_2)]_+\big\rangle_{\gamma_i=0} , \nonumber\\
	g_{ik}^{\rm R}(t_1,t_2) &= i  \big\langle c_{ik}^{\dagger}(t_2) c_{ik}(t_1) \big\rangle_{\gamma_i=0} . 
\end{align}
The dynamic conductance in Eq.~(\ref{Eq:dyn-cond})  is  
\begin{align}
	&\mathcal{G}_{i}^{j}(t,t_1) 
	= \frac{e}{2} \Bigl\{ G^{\rm R} \big| \Sigma_{j}^{\rm R} G^{\rm R} \Sigma_i^< + G^{\rm R} \Sigma_{j}^{\rm R} \big| G^{\rm R} \Sigma_i^<  \nonumber \\
	&\hspace{1.0cm} - \Sigma_i^< G^A \big| \Sigma_{j}^A G^A  + \Sigma_i^< G^A  \Sigma_{j}^A \big| G^A  \nonumber \\
	&\hspace{1.0cm} +  G^{\rm R} \big| \Sigma_{j}^{\rm R} G^< \Sigma_i^A +  G^{\rm R} \Sigma_{j}^{\rm R} \big| G^< \Sigma_i^A + G^{\rm R} \big| \Sigma_{j}^< G^A \Sigma_i^A \nonumber \\
	&\hspace{1.0cm} +  G^{\rm R} \Sigma_{j}^< \big| G^A \Sigma_i^A + G^< \big| \Sigma_{j}^A G^A \Sigma_i^A + G^< \Sigma_{j}^A \big| G^A \Sigma_i^A \nonumber \\
	&\hspace{1.0cm}-  \Sigma_i^{\rm R} G^{\rm R} \big| \Sigma_{j}^{\rm R} G^< - \Sigma_i^{\rm R} G^{\rm R} \Sigma_{j}^{\rm R} \big| G^<  - \Sigma_i^{\rm R} G^{\rm R} \big| \Sigma_{j}^< G^A  \nonumber \\
	&\hspace{1.0cm} - \Sigma_i^{\rm R} G^{\rm R} \Sigma_{j}^< \big| G^A  - \Sigma_i^{\rm R} G^< \big| \Sigma_{j}^A G^A  - \Sigma_i^{\rm R} G^< \Sigma_{j}^A \big| G^A  \nonumber \\
	&\hspace{1.0cm}+ \delta_{i,j} (G^{\rm R} \big| \Sigma_i^<   + G^< \big| \Sigma_i^A  - \Sigma_i^{\rm R} \big| G^<  - \Sigma_i^< \big| G^A) \Bigr\} ,
	\nonumber
\end{align}
where we define $AB \big| CD = [AB](t,t_1) [CD](t_1,t)$
with $[AB](t_1,t_2) = \int dt_3 \ A(t_1,t_3) B(t_3,t_2)$.

One can Fourier transform $G^{A/R}(t_1,t_2)$ to get 
\begin{eqnarray}
G^{\rm A}(\omega) = \big[G^{\rm R}(\omega)\big]^* =  \Big(\omega-\epsilon_{\rm d} -\Lambda(\omega) + \rmi {\textstyle {1\over 2}} \Gamma\Big)^{-1}. \qquad
\label{Eq:GA}
\end{eqnarray}

%==================================================

\bibliography{references}

%============================================
%============================================
%============================================

\end{document}